\documentclass[conference,a4paper]{IEEEtran}
\usepackage[left=1.43cm,right=1.43cm,top=1.82cm,bottom=4.3cm]{geometry}

\IEEEoverridecommandlockouts
\usepackage{graphicx}
\usepackage{amssymb}
\usepackage{mathtools}
\usepackage{dsfont}
\usepackage{cite}
\usepackage{stfloats}
\usepackage{psfrag}
\usepackage[mathscr]{euscript}
\usepackage{acronym}  % make an acronym
\usepackage{amsmath}
\usepackage{algorithm}
\usepackage[noend]{algpseudocode}
\usepackage{booktabs}
\usepackage{bbm}
\usepackage{subcaption}
\usepackage{graphicx}
\usepackage{caption} 
\usepackage{comment}

\makeatletter
\def\BState{\State\hskip-\ALG@thistlm}
\makeatother

\usepackage{float}
\usepackage{balance}

\acrodef{CCDF}{complementary cumulative distribution function}
\acrodef{CF}{characteristic function}
\acrodef{PPP}{Poisson point processe}
\acrodef{RV}{random variable}
\acrodef{i.i.d.}{independent and identically distributed}
\acrodef{PDF}{probability distribution function}
\acrodef{CDF}{cumulative distribution function}
\acrodef{ch.f.}{characteristic function}
\acrodef{AWGN}{additive white Gaussian noise}
\acrodef{SNR}{signal-to-noise ratio}
\acrodef{LRT}{likelihood ratio test}
\acrodef{DRT}{distance ratio test}
\acrodef{GLRT}{generalized likelihood ratio test}
\acrodef{CRLB}{Cram\'{e}r-Rao lower bound}
\acrodef{CRB}{Cram\'{e}r-Rao bound}
\acrodef{ZZLB}{Ziv-Zakai lower bound}
\acrodef{ZZB}{Ziv-Zakai bound}
\acrodef{LOS}{line-of-sight}
\acrodef{ToF}{time-of-flight}
\acrodef{NLOS}{non-line-of-sight}
\acrodef{GDOP}{geometric dilution of precision}
\acrodef{GPS}{Global Positioning System}
\acrodef{FIM}{Fisher information matrix}
\acrodef{PEB}{position error bound}
\acrodef{SPEB}{squared position error bound}
\acrodef{TOA}{time-of-arrival}
\acrodef{TOF}{time-of-flight}
\acrodef{WSN}{wireless sensor network}
\acrodef{MAC}{medium access control}
\acrodef{RSS}{received signal strength}
\acrodef{WAF}{wall attenuation factor}
\acrodef{TDOA}{time difference-of-arrival}
\acrodef{RF}{radiofrequency}
\acrodef{RTT}{round-trip time}
\acrodef{AOA}{angle-of-arrival}
\acrodef{MF}{matched filter}
\acrodef{ED}{energy detector}
\acrodef{ML}{maximum likelihood}
\acrodef{MSE}{mean-square error}
\acrodef{RMSE}{root-mean-square error}
\acrodef{LEO}{localization error outage}
\acrodef{ppm}{part-per-million}
\acrodef{ACK}{acknowledge}
\acrodef{UWB}{Ultrawide bandwidth}
\acrodef{TNR}{threshold-to-noise ratio}
\acrodef{LS}{least squares}
\acrodef{IR-UWB}{impulse radio UWB}
\acrodef{FCC}{Federal Communications Commission}
\acrodef{TH}{time-hopping}
\acrodef{PPM}{pulse position modulation}
\acrodef{MUI}{multi-user interference}
\acrodef{PDP}{power delay profile}
\acrodef{BPZF}{band-pass zonal filter}
\acrodef{SIR}{signal-to-interference ratio}
\acrodef{SINR}{signal-to-interference-plus-noise ratio}
\acrodef{RFID}{radio frequency identification}
\acrodef{WPAN}{wireless personal area network}
\acrodef{WWB}{Weiss-Weinstein bound}
\acrodef{DP}{direct path}
\acrodef{MF}{matched filter}
\acrodef{MMSE}{minimum-mean-square-error}
\acrodef{SBS}{serial backward search}
\acrodef{SBSMC}{serial backward search for multiple clusters}
\acrodef{NBI}{narrowband interference}
\acrodef{WBI}{wideband interference}
\acrodef{INR}{interference-to-noise ratio}
\acrodef{CR}{channel response}
\acrodef{CIR}{channel impulse response}
\acrodef{CR}{channel  response}
\acrodef{RADAR}{radar}
\acrodef{MUR}{Multistatic radar}
\acrodef{JBSF}{jump back and search forward}
\acrodef{HDSA}{high-definition situation-aware}
\acrodef{RRC}{root raised cosine}
\acrodef{ST}{simple thresholding}
\acrodef{BTB}{Bellini-Tartara bound}
\acrodef{P-Max}{$P$-Max}  %suggestion, use with \acl{P-Max}
\acrodef{MIMO}{multiple-input multiple-output}
\acrodef{MAP}{maximum a posteriori}
\acrodef{FG}{factor graph}
\acrodef{OP}{outage probability}
\acrodef{WED}{wall extra delay}
\acrodef{RMS}{root mean square}
\acrodef{SPAWN}{sum-product algorithm over a wireless network}
\acrodef{MDD}{minimum distance distribution}
\acrodef{MAP}{maximum a posteriori probability}
\acrodef{SAP}{small cell access point}
\acrodef{UE}{user equipment}
\acrodef{MBS}{macro cell base station}
\acrodef{UER}{\ac{UE} Relay}
\acrodef{D2D}{device-to-device}
\acrodef{MBS}{macro base station}
\acrodef{CSI}{channel state information}
\acrodef{OGR}{outage guard region}
\acrodef{FUR}{feasible UER region}
\acrodef{EHR}{energy harvesting region}
\acrodef{EH}{energy harvesting}
\acrodef{D2D-EHSN}{D2D communication provided \ac{EH} small cell network}
\acrodef{D2D-EHHN}{D2D communication provided \ac{EH} heterogeneous network}
\acrodef{3GPP}{3rd Generation Partnership Project}
\acrodef{BS}{base station}
\acrodef{DF}{decode and forward}
\acrodef{CCDF}{complementary cumulative distribution function}
\acrodef{ZF}{zero forcing}
\acrodef{RZF}{regularized zero forcing}
\acrodef{WLLN}{weak law of large number}
\acrodef{SLLN}{strong law of large numbers}
\acrodef{TDD}{Time-division duplex}
\acrodef{EE}{energy efficiency} 
\acrodef{HetNet}{heterogeneous network} 
\acrodef{SCP}{Single Cell Processing}
\acrodef{CBF}{Coordinated Beamforming}
\usepackage{color}
\usepackage{dsfont}
\usepackage{bbm}

\DeclareMathAlphabet{\mathsf}{OML}{cmbr}{m}{it}

\newtheorem{theorem}{\bf Theorem}
\newtheorem{lemma}{\bf Lemma}

\newcommand{\bd}{\begin{description}}
\newcommand{\ed}{\end{description}}
\newcommand{\be}{\begin{enumerate}}
\newcommand{\ee}{\end{enumerate}}
\newcommand{\bi}{\begin{itemize}}
\newcommand{\ei}{\end{itemize}}
\newcommand{\bl}{\begin{list}}
\newcommand{\el}{\end{list}}
\newcommand{\bt}{\begin{tabbing}}
\newcommand{\et}{\end{tabbing}}

\newcommand{\paperTitle}{The Meta Distribution of the SIR in Joint Communication and Sensing Networks }

\begin{document}

% This code is to reduce the list of authors by using et. al:
\bstctlcite{IEEEexample:BSTcontrol}

{
\title{\paperTitle}

\author{
\IEEEauthorblockN{Kun~Ma$^*$, Chenyuan Feng$^\dagger$, Giovanni Geraci$^\sharp$, and Howard~H.~Yang$^*$
\vspace{0.1cm}} 
\IEEEauthorblockA{$^*$ \textit{ZJU-UIUC Institute, Zhejiang University, Haining, China}}
\IEEEauthorblockA{$^\dagger$ \textit{Department of Communication, EURECOM, France}}
\IEEEauthorblockA{$^{\sharp}$ \textit{Telef\'{o}nica Research and Universitat Pompeu Fabra, Barcelona, Spain}}

\thanks{K. Ma and H. H. Yang were supported by the National Natural Science Foundation of China under Grant 62201504, by the Zhejiang Provincial Natural Science Foundation of China under Grant LGJ22F010001, and by the Zhejiang Lab Open Research Project (No. K2022PD0AB05).
The work of C. Feng was partially carried out in the context of Beyond5G, a project funded by the French government as part of the economic recovery plan, namely "France Relance", and the investments for the future program.
G. Geraci was supported by the Spanish Research Agency grants PID2021-123999OB-I00 and CEX2021-001195-M, by the UPF-Fractus Chair, and by the Spanish Ministry of Economic Affairs and Digital Transformation and the European Union NextGenerationEU through the UNICO 5G I+D SORUS project.}
}

\maketitle
\acresetall
\thispagestyle{empty}

\begin{abstract}
In this paper, we introduce a novel mathematical framework for assessing the performance of joint communication and sensing (JCAS) in wireless networks, employing stochastic geometry as an analytical tool. We focus on deriving the meta distribution of the signal-to-interference ratio (SIR) for JCAS networks. This approach enables a fine-grained quantification of individual user or radar performance intrinsic to these networks. 
Our work involves the modeling of JCAS networks and the derivation of mathematical expressions for the JCAS SIR meta distribution. Through simulations, we validate both our theoretical analysis and illustrate how the JCAS SIR meta distribution varies with the network deployment density.
\end{abstract}

% \begin{IEEEkeywords}
% Joint communication and sensing, stochastic geometry, meta SIR distribution, coverage probability. \red{Gio: not needed for conference. Save space.} 
% \end{IEEEkeywords}

\IEEEpeerreviewmaketitle

%%%%%%%%%%%%%%%%%%%%%%%%%%%%%%%%%%%%%%%%%%%%%%%%%%%%
\section{Introduction}\label{sec:intro}

One of the envisioned features of sixth generation (6G) mobile networks is the synergy between wireless communications and sensing \cite{TarKhaWon:20WCMAG,OugGerPol2023,DanAmiShi:20NE}. Joint communication and sensing (JCAS) paves the way for diverse applications, such as indoor localization, autonomous aircraft, and extended reality \cite{liu2020joint,zhang2021enabling}, while also presenting a challenge in efficiently sharing the spectrum. To address this, a detailed analysis of the underlying trade-offs is necessary, with well-defined performance metrics to gauge the network's effectiveness.

Stochastic geometry emerges as a robust analytical framework for this task, offering theoretical models to characterize network-wide performance. For instance, time-sharing networks were explored in \cite{ren2020performance}, detailing their radar detection range, false alarm rates, and communication success probabilities. Other studies have focused on the energy and spectrum efficiency in base station (BS) deployments \cite{salem2022rethinking} and the extension of coverage probability and ergodic capacity concepts to radar settings \cite{olson2022coverage}. Notably, JCAS techniques are also crucial in vehicular networks, with studies investigating the detection range in spectrum-sharing scenarios and the impact of interference on road obstacle detection and communication \cite{ghozlani2021stochastic,ma2022performance}.

Despite these advancements, a gap remains in understanding the individual performance of users or radars within a JCAS network. Existing studies have primarily concentrated on the coverage probability, a geographic average that only provides information about the expected JCAS performance across all network deployments, overlooking the variability in user or radar experiences. For example, while one network realization may exhibit a wide range of success probabilities (e.g., 0.5 to 0.99), another might show a narrower range (e.g., 0.85 to 0.95), yet both could have the same spatial average. %This points to a need for a more fine-grained performance analysis tool.

%%%
In this paper, we employ a more fine-grained tool and analyze the performance of JCAS networks by deriving the meta distribution of the signal-to-interference ratio (SIR), a concept previously introduced in \cite{haenggi2016meta}. While the SIR meta distribution has been applied separately to either communication or radar detection scenarios \cite{Hae:21CL-1,Hae:21CL-2,ghatak2022radar,ram2022optimization}, our work uniquely applies it to JCAS networks. Our contributions can be summarized as follows:
\begin{itemize}
  \item 
  We establish an analytical framework for modeling the JCAS network, based on which we derive exact mathematical expressions for the conditional JCAS coverage probability and (the complementary of) its distribution, namely, the SIR meta distribution.  
  % %
  % \item 
  % As the above exact expressions may require a considerable effort to be numerically evaluated, we also propose tight approximations based on practical assumptions.
  %
  \item 
  We validate our analysis through simulations and present numerical results to illustrate the behavior of the JCAS SIR meta distribution with respect to the network deployment density.
\end{itemize}
%%%%%%%%%%%%%%%%%%%%%%%%%%%%%%%%%%%%%%%%%%%%%%%%%%%%
\section{System Model}\label{sec:sysmod}
\subsection{Network Deployment}

%%%%%%%%%%%%%%%%%%%%%%%%%%%%%%%%%%%%%%%%%%%%%%%%%%%%
\begin{figure}[t!]
  \centering{}
    {\includegraphics[width=0.9\columnwidth]{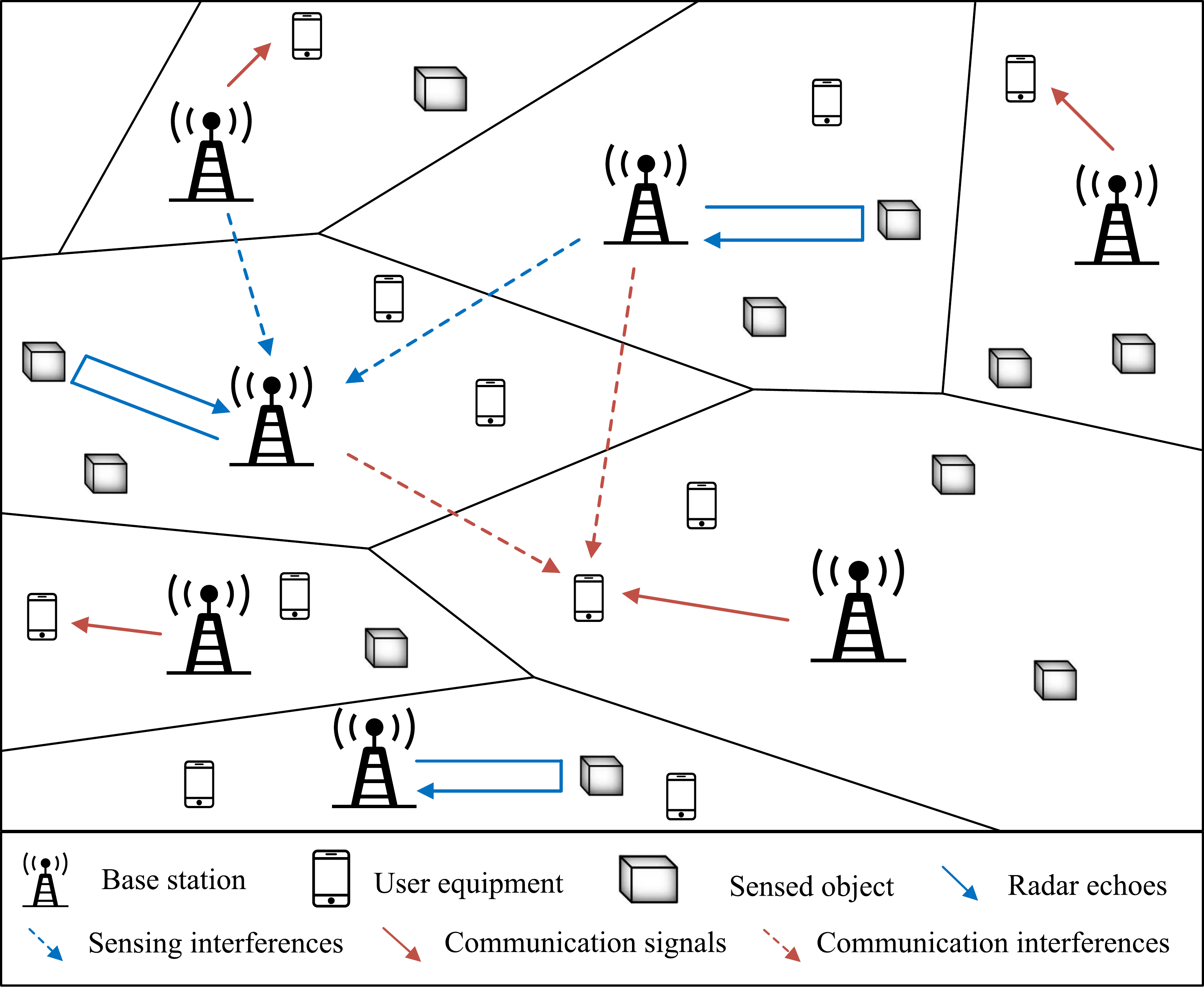}}
  \caption{Illustration of the JCAS network considered, with BSs simultaneously sending information packets to UEs and sensing waveforms to the SOs for which they receive radar echoes.}
  \label{fig:system model}
\end{figure}
%%%%%%%%%%%%%%%%%%%%%%%%%%%%%%%%%%%%%%%%%%%%%%%%%%%%

We consider the JCAS wireless network depicted in Fig.~\ref{fig:system model}, consisting of multiple base stations (BSs), users (UEs), and sensed objects (SOs). 
The locations of the BSs, UEs, and SOs are modeled as three independent homogeneous Poisson point processes (PPPs), denoted by $\Phi_\mathrm{b}$, $\Phi_\mathrm{u}$, and $\Phi_\mathrm{s}$ with intensities of $\lambda_\mathrm{b}$, $\lambda_\mathrm{u}$, and $\lambda_\mathrm{s}$, respectively.
The BSs are in charge of sending information packets to UEs in the downlink or sending sensing waveforms to the SOs and receiving echoes from them.
We assume communication and sensing to occur in separate time-frequency resources and synchronously for all BSs. 
%We consider that communication and sensing functionalities are carried out via 
A shared multicarrier waveform is universally reused across the network; hence, the transmissions of different nodes will interfere with each other.

All of the BSs are assumed to be active, where each BS selects one entity (which can be a UE or SO) in its coverage uniformly at random, and performs the corresponding service. 
We assume $\lambda_\mathrm{u} \gg \lambda_\mathrm{b} $ and $\lambda_\mathrm{s} \gg \lambda_\mathrm{b}$. As such, every cell would contain multiple UEs and SOs.  
We further assume that any pair of wireless transmission channels is subject to path loss, which obeys the power law, and Rayleigh fading, with the BSs transmitting at a fixed power $P_{\mathrm{tx} }$.
% By utilizing the maximum average received power association policy, the UEs and SOs associate with the nearest BSs in space. 

\subsection{SIR Models}
We first characterize the large-scale path loss of the propagation model. 
The link between a pair of transmitter and receiver can be line-of-sight (LoS) or non-line-of-sight (NLoS), depending on whether blockage intersects the link. 
For a link of length $r$, the LoS probability is given by $p_\mathrm{L}(r) = \exp(-\beta r)$ \cite{bai2014analysis}, and the NLoS probability is $p_\mathrm{N}(r) = 1-p_\mathrm{L}(r)$, where $\beta$ is a parameter capturing the density of blockages. 
Note that communications can occur over a LoS or NLoS link, characterized by different system factors. 
Specifically, the path loss function with the length $r$ is given by
\begin{align} \label{equ:PL_c}
	L_1(r) = 
 \begin{cases}
K_\mathrm{L} r^{-\alpha_\mathrm{L}}, \quad \text{w.\,p.} \quad p_\mathrm{L}(r) \\
K_\mathrm{N} r^{-\alpha_\mathrm{N}}, \quad \text{w.\,p.} \quad 1 - p_\mathrm{L}(r)
\end{cases}    
\end{align}
where $K_\mathrm{L}$ and $K_\mathrm{N}$ are channel gains associated with the LoS and NLoS links, respectively, while $\alpha_\mathrm{L}$ and $\alpha_\mathrm{N}$ are path loss exponents of the LoS and NLoS links, respectively.

Based on Slivnyark's theorem \cite{Zuyev2006}, we can concentrate on a typical UE located at the origin. 
The typical UE is associated with the BS, either LoS or NLoS, with the smallest path loss.
Let $X_0$ denote the location of the tagged BS (also referred to as the \textit{typical BS}) of the \textit{typical UE} and $X_k$ denote the location of the $k$-th BS in the network.
Then, the SIR received at the typical UE can be expressed as  
\begin{align} \label{equ:SIRc}
	\mathrm{SIR_{c}} = \frac{ h_{0} L_1(\Vert X_{0} \Vert) }{ \sum_{ k\neq 0 }  h_{k} L_1(\Vert X_k \Vert)},
\end{align}
where $h_{k} \sim \exp(1)$ is the channel fading from the $k$-th BS to the typical UE and $\Vert \cdot \Vert$ denotes the Euclidean norm. 

Similarly, we consider a monostatic sensing scenario where the BSs transmit the sensing waveforms and listen to the echo of the SOs. We assume sensing links are LoS only, because detecting or estimating the location of a NLoS object is highly challenging in practice. 
Hence, the path loss function of a monostatic radar can be derived as 
\begin{align} \label{equ:PL_s}
	L_2(r) = 
K_\mathrm{L} r^{-2\alpha_\mathrm{L}}.   
\end{align}

Without loss of generality, we place the \textit{typical SO} at the origin \cite{olson2022coverage}. 
Notably, the typical SO is associated with the nearest BS with a LoS path, denoted as $X_0$. As such, the signal strength of the radar echo measured at the typical BS is given by \cite{richards2005fundamentals}
\begin{align}
	S_{\mathrm{s}} =\frac{P_{\mathrm{tx} } G_t G_r \lambda_{ \mathrm{w} } ^2 \sigma_{ \mathrm{cs} } L_2 (\Vert X_{0} \Vert )}{(4\pi)^3}=	A\frac{\sigma_{ \mathrm{cs} }}{4\pi}L_2(\Vert X_{0} \Vert),
\end{align}
where $G_t$ and $G_r$ denote the antenna gains of transmission and reception in the sensing stage, respectively, $\lambda_{ \mathrm{w} }$ is the carrier wavelength, $\sigma_{ \mathrm{cs} } \sim \exp(1)$ represents the radar cross-section, which can be modeled as a random variable that follows the exponential distribution with unit mean \cite{shnidman2003expanded}, and $A = P G_t G_r (\lambda_{ \mathrm{w} } /4\pi)^2 $.

Since sensing employs the same carrier as communication, the interference accumulated at the typical BS from all other BSs can be expressed as 
\begin{align}
	I_{\mathrm{s}} =\sum_{ k \neq 0 }A \tilde{h}_{k} L_1(\Vert  X_k-X_0 \Vert),
\label{eqn:Is}
\end{align}
where $\tilde{h}_{k} \sim \exp(1)$ stands for the channel fading from the $k$-th BS to the typical SO. We neglect the interference caused by other radar echoes due to its lower power compared to (\ref{eqn:Is}).
% As a simplification, we assume the LoS/NLoS status of the interfering BSs with respect to the serving BS are independent of the LoS/NLoS status of the BSs observed by the typical SO.
By analogy with the communication scenario, the sensing SIR at the average BS can then be written as follows 
\begin{align} \label{equ:SIRs}
	\mathrm{SIR_{ \mathrm{s} }} &=\mathbbm{1} \{\Vert X_{0} \Vert < \infty\} \frac{S_{\mathrm{s}}}{I_{\mathrm{s}}} \nonumber \\ &= \mathbbm{1}\{\Vert X_{0} \Vert < \infty\} \frac{ \frac{ \sigma_{ \mathrm{cs} } }{ 4\pi } L_2(\Vert X_{0} \Vert) }{ \sum_{ k \neq 0 }  \tilde{h}_{k}L_1(\Vert X_k-X_0 \Vert)}
\end{align}
%I move this part to the previous paragraph
%\green{where we neglect the interference caused by other radar echoes}
where $\mathbbm{1}\{\Vert X_{0} \Vert < \infty\}$ indicates that sensing is LoS only, and thus, SIR is zero if there are no LoS BSs to the typical SO.

We note that while $\mathrm{SIR_{c}}$ is an actual SIR, $\mathrm{SIR_{ \mathrm{s} }}$ is a conceptual one, constructed as a proxy for the BS's efficacy in estimating the SO's parameter of interest.{\footnote{In the following, for the sake of readability, we neglect the constant multiplier $\frac{1}{4\pi}$ in \eqref{equ:SIRs} as it can be embedded into the decoding threshold.}}
Based on these SIR models, we can establish suitable metrics to assess the JCAS network performance.

\subsection{Performance Metric}

The probability that $\mathrm{SIR_c}$ surpasses a decoding threshold $\theta_{\mathrm{c}}$ is known as the coverage probability, a widely used metric to evaluate link performance in cellular networks. 
This metric provides information about the fraction of UEs in the network that achieves a SIR at least at the level of $\theta_{\mathrm{c}}$. 
The sensing performance can be defined in a similar manner. 
The estimation rate, which is the mutual information between the radar return and the parameter of interest divided by coherent processing interval, can be used to characterize the quality of the sensing, with upper and lower bounds determined by logarithmic functions of $\mathrm{SIR_s}$ \cite{olson2022coverage}. 
Hence, the sensing accuracy can be captured using a measure based on the distribution of $\mathrm{SIR_s}$. For instance, one could consider the sensing coverage probability, defined as the probability of $\mathrm{SIR_s}$ surpassing a predetermined threshold $\theta_s$ and reflecting the average portion of the SOs whose SIR reaches $\theta_\mathrm{s}$. 

Since the coverage probabilities only provide information about the average JCAS performance across all network deployments, in this paper, we leverage the notion of conditional coverage probability and meta SIR distribution \cite{haenggi2016meta,Hae:21CL-1,Hae:21CL-2} to obtain a fine-grained perspective of the network performance. Specifically, given the point process $\Phi_\mathrm{b}$, we define the \emph{conditional JCAS coverage probability} as the joint fraction of UEs or SOs whose corresponding SIR is above their corresponding threshold, given by
\begin{align}\label{equ:Cnd_JCAS_cvrg}
	P( \theta_\mathrm{c}, \theta_\mathrm{s} ) &= \mathbb{P}_{\Phi_{\mathrm{u}} + \Phi_{\mathrm{s}}}(\mathrm{SIR} > \theta \mid \Phi_{\mathrm{b}}) \nonumber \\
 & \stackrel{(a)}=\frac{\lambda_\mathrm{u}}{\lambda_\mathrm{u} + \lambda_\mathrm{s}} P_\mathrm{c}(\theta_\mathrm{c}) +\frac{\lambda_\mathrm{s}}{\lambda_\mathrm{u} + \lambda_\mathrm{s}}P_\mathrm{s}(\theta_\mathrm{s})
\end{align}
where $P_\mathrm{c}(\theta_\mathrm{c}) = \mathbb{P}_{\Phi_{\mathrm{u}}}(\mathrm{SIR}_\mathrm{c} > \theta_\mathrm{c} \mid \Phi_{\mathrm{b}} ), P_\mathrm{s}(\theta_\mathrm{s}) = \mathbb{P}_{\Phi_{\mathrm{s}}}(\mathrm{SIR}_\mathrm{s} > \theta_\mathrm{s} \mid \Phi_{\mathrm{b}} ) $ stand for the conditional communication and sensing coverage probabilities, respectively, and ($a$) follows from the independence of $\Phi_{\mathrm{u}}$ and $\Phi_{\mathrm{s}}$ and using the superposition theorem for the stationary process.

We note that $P( \theta_\mathrm{c}, \theta_\mathrm{s})$ is still a random variable, since while channel fading is averaged out, the randomness stemming from $\Phi_\mathrm{b}$ remains.%
\footnote{The conventional coverage probability can be obtained by taking the expectation of $P( \theta_\mathrm{c}, \theta_\mathrm{s} )$ with respect to $\Phi_\mathrm{b}$, thereby disregarding the dependence of the JCAS performance on the network realization $\Phi_\mathrm{b}$.}
In that respect, we leverage the concept of the SIR meta distribution in communication networks \cite{haenggi2016meta} and define the \emph{JCAS SIR meta distribution} as the complementary cumulative distribution function (CCDF) of  $P( \theta_\mathrm{c}, \theta_\mathrm{s} )$, i.e., 
\begin{align} \label{equ:Meta_JCAS}
&F(\theta_\mathrm{c}, \theta_\mathrm{s},x) = \mathbb{P}( P( \theta_\mathrm{c}, \theta_\mathrm{s} ) > x ). 
\end{align}
This quantity provides information about the fraction of end terminals (UEs or SOs) in the network that can attain the desired SIR (at the level of $\theta_{ \mathrm{c} }$ and $\theta_{ \theta{s} }$ for  $\mathrm{SIR}_{ \mathrm{c} }$ and $\mathrm{SIR}_{ \theta{s} }$, respectively) with reliability (i.e., probability) of at least $x$.

% The meta distribution $F(\theta_\mathrm{c}, \theta_\mathrm{s},x)$ quantifies the joint fraction of UEs and SOs in the network that can attain $\mathrm{SIR}_{c}$ of $\theta_{\mathrm{c}}$ or $\mathrm{SIR}_{s}$ of $\theta_s$, with a  reliability (i.e., probability) of at least $x$. 

%========================================%
%            Analysis
%========================================%
\section{Analysis of JCAS SIR Meta Distribution} \label{Sec:Analysis}

This section details the steps to derive analytical expressions for the quantity in \eqref{equ:Meta_JCAS}.
We first calculate the general moments of conditional sensing and communication coverage probabilities.
Then, we derive the analytical expression of JCAS SIR meta distribution. 
% We specifically focus on the sensing case since communication case have been well studied.
% According  to \eqref{equ:Meta_JCAS}, the JCAS SIR meta distribution can by obtained by characterizing the communication and sensing SIR meta distribution separately. 
% \subsection{Analysis of the JCAS SIR Meta Distribution}

\subsubsection*{Conditional sensing coverage probability}
We begin by deriving the conditional sensing coverage probability $P_\mathrm{s}( \theta_\mathrm{s})$ by averaging out the randomness in channel fading. The quantity $P_\mathrm{s}( \theta_\mathrm{s})$ represents the probability that, given a network realization $\Phi_{ \mathrm{b} }$, the effect of channel fading results in sensing $\mathrm{SIR}_\mathrm{s}$ exceeding $\theta_{ \mathrm{s}}$. 
        
\begin{lemma} \label{lma:Cnd_sensing_cvrg}
	\textit{ 
 Conditioned on the point process $\Phi_{ \mathrm{b} }$, the sensing coverage probability is given by
		\begin{align}\label{equ:cnd_cvrg_exact}
			P_\mathrm{s}( \theta_\mathrm{s} ) = \Big( 1 - e^{ - \frac{ 2\pi\lambda_{ \mathrm{b} }}{\beta^2} } \Big) \!\!\!\! \prod_{i \in \{ \mathrm{L},\mathrm{N} \} } \prod_{ X_k \in \Phi^i_{ \mathrm{b} \setminus X_0 }  } \!\!\!\! \Bigg(  \frac{1}{1 + \frac{ \theta_\mathrm{s} K_i\Vert X_0 \Vert ^{\alpha_\mathrm{L}}}{ K_{\mathrm{L}}\Vert X_k -X_0 \Vert^{\alpha_i}} } \Bigg) 
   % \nonumber 
   %  \\ &\times  
   %  \Big( 1-\exp\Big(\frac{-2\pi\lambda_{ \mathrm{b} }}{\beta^2}\Big)\Big)
		\end{align}
  where $\Phi^{ \mathrm{L} }_{ \mathrm{b} \setminus X_0 }$ and $\Phi^{ \mathrm{N} }_{ \mathrm{b} \setminus X_0 }$ represent the interfering nodes in $\Phi_{ \mathrm{b} }$ with LoS and NLoS paths to the serving BS, respectively. 
	}
\end{lemma}
\begin{IEEEproof}
    Using \eqref{equ:SIRs}, we can calculate the conditional JCAS coverage probability as 
	\begin{align}
    &P( \theta_\mathrm{s} )  = \mathbb{P}_{\Phi_\mathrm{s}}(\mathrm{SIR}_\mathrm{s} > \theta_\mathrm{s} \mid \Phi_\mathrm{b} ) \nonumber \\
   &\stackrel{(a)} = \mathbb{P}\Big( \Vert X_{0} \Vert < \infty, \Phi_{  \mathrm{b} } \Big) \nonumber \\
   &\quad~ \times \mathbb{P} \Big( \sigma_{\mathrm{cs}} >
    \frac{ \theta_{ \mathrm{s} } \sum_{ k \neq 0 }  \tilde{h}_{k} L_1( \Vert X_k-X_0 \Vert) }{L_2(\Vert X_0 \Vert)}  \mid \Vert X_{0} \Vert < \infty, \Phi_{  \mathrm{b} } \Big) 
    \nonumber \\ 
    % &\times  \Big( 1- \exp\Big(\frac{-2\pi\lambda_{ \mathrm{b} }}{\beta^2} \Big) \Big) \nonumber \\
    &\stackrel{(b)} = \mathbb{E} \Big[ \exp\Big( \frac{-\theta_{ \mathrm{s} }}{ L_2(\Vert X_{0} \Vert )}\sum_{ k \neq 0 }  \tilde{h}_{k}L_1(\Vert X_k-X_0 \Vert) \Big) \mid  \Vert X_{0} \Vert < \infty \Big]\nonumber 
    \\ &  
    \qquad \qquad \qquad \qquad \qquad \qquad \qquad \quad \times \Big( 1- \exp\Big(\frac{-2\pi\lambda_{ \mathrm{b} }}{\beta^2} \Big) \Big) \nonumber \\
    &\stackrel{(c)} = \prod_{i \in \{ \mathrm{L},\mathrm{N} \} } \prod_{ X_k \in  \Phi^i_{ \mathrm{b}\setminus X_0 }   } \Bigg(  \frac{1}{1 + \theta_\mathrm{s}\frac{ K_i\Vert X_0 \Vert ^{\alpha_\mathrm{L}}}{ K_{\mathrm{L}}\Vert X_k - X_0\Vert^{\alpha_i}} } \Bigg)  
    \nonumber  \\ 
    &\qquad \qquad \qquad \qquad \qquad \qquad \qquad \quad \times 
    \Big( 1-\exp\Big(\frac{-2\pi\lambda_{ \mathrm{b} }}{\beta^2}\Big)\Big)
     %&\stackrel{(c)} = \Big( 1 - e^{ - \frac{ 2\pi\lambda_{ \mathrm{b} }}{\beta^2} } \Big) \prod_{i \in \{ \mathrm{L},\mathrm{N} \} } \prod_{ X_k \in  \Phi^i_{ \mathrm{b} } \setminus X_0  } \Big(  \frac{1}{1 + \theta_\mathrm{s}\frac{ K_i\Vert X_0 \Vert ^{\alpha_\mathrm{L}}}{ K_{\mathrm{L}}\Vert X_k - X_0\Vert^{\alpha_i}} } \Big) 
	\end{align}
where ($a$) follows from that $P(\Vert X_{0} \Vert < \infty) = 1-\exp(\frac{-2\pi\lambda_{ \mathrm{b} }}{\beta^2})$\cite{bai2014coverage}, ($b$) follows because $\sigma_{ \mathrm{cs} }$ obeys exponential distributions, while ($c$) holds since fading realizations are also exponentially distributed and mutually independent and the independence of point processes $\Phi^{ \mathrm{L} }_{ \mathrm{b} \setminus X_0 }$ and $\Phi^{ \mathrm{N} }_{ \mathrm{b} \setminus X_0 }$.
\end{IEEEproof}

\subsubsection*{Moment of conditional sensing coverage probability}
Next, we derive the expression of the moment of $P(\theta_\mathrm{s} )$.

\begin{theorem} \label{thrm:momts_Sen_cvrg}
	\textit{
 % The meta distribution of the SIR in the sensing scenario under consideration is given by 
	% 	\begin{align}\label{equ:meta}
	% 		F( \theta_\mathrm{s}, x) = \frac{1}{2} - \frac{1}{\pi} \int_{0}^{\infty} \mathrm{Im} \Big\{ x^{ - j \omega } M_{j\omega}^\mathrm{s} \Big\} \frac{d \omega}{ \omega },
	% 	\end{align}
	% 	 where $j = \sqrt{-1}$, $\mathrm{Im}\{ \cdot \}$ denotes the imaginary part of the input variable, and 
  The $b$-th moment of conditional sensing coverage probability $P_\mathrm{s}(\theta_\mathrm{s})$ is given by
		\begin{align}\label{equ:M_b}
			M_b^\mathrm{s} = &\Big( 1-\exp\Big( \frac{-2\pi \lambda_{ \mathrm{b} }}{\beta^2} \Big) \Big)^{b-1} 
 \int_{0}^{\infty} f(r_0) \exp\Big(-\sum_{i\in \{\mathrm{L},\mathrm{N}\}}\nonumber \\
 \int_{0}^{\infty}&\Big(1 - \frac{1}{(1 + \theta_\mathrm{s}\frac{ K_i}{ K_{\mathrm{L}}} r_0^{2\alpha_\mathrm{L}}r^{-\alpha_i})^b} \Big) \lambda_i(r,r_0)dr\Big) dr_0, 
		\end{align}
		in which 
            \begin{align}
                f(r_0) = 2\pi \lambda_{ \mathrm{b} } r_0p_{\mathrm{L}}(r_0)\exp\Big(\frac{-2\pi\lambda_{ \mathrm{b} }}{\beta^2}(1-e^{-\beta r_0}(\beta r_0 + 1)) \Big),
            \end{align}
            and
            \begin{align}
                \lambda_i(r,r_0)=2p_i(r)\lambda_{ \mathrm{b} } r\Big(\pi-J(r,r_0)\mathbbm{1}\{r \le 2r_0\}\Big),
            \end{align}
            where
            \begin{align}\label{equ:Jr}
                J(r,r_0) =\! \int_{\frac{r}{2r_0}}^1 \! (1-u^2)^{-\frac{1}{2} }\exp\!\Big(\! -\beta \sqrt{r^2-2rr_0u+r_0^2}\Big)du.
            \end{align}
	}
\end{theorem}
\begin{IEEEproof}
Please refer to Appendix A.
\end{IEEEproof}

\subsubsection*{Moment of conditional communication coverage probability}
Following the similar manner to that of Lemma 1, given that the typical UE is associated with a LoS/NLoS BS, the conditional communication coverage probability can be expressed as
\begin{align}
    P_\mathrm{c,\rho}(\theta_\mathrm{c}) =\!\!\!\!\!\! \prod_{i\in \{\mathrm{L},\mathrm{N}\}} \prod_{ X_k \in \Phi_\mathrm{i} }  \!\!\!\! \Bigg(
    \frac{1}{1 \!+\! \frac{\theta_\mathrm{c} K_i \Vert X_0 \Vert ^{ \alpha_\rho }}{K_\rho \Vert X_k \Vert ^{\alpha_i}}}
    \Bigg), \rho \in \{ \mathrm{L},\mathrm{N}\}.
\end{align}
where $\rho$ denotes the LoS/NLoS status of serving BS and $\Phi_\mathrm{L}$/$\Phi_\mathrm{N}$ denotes the interfering BSs whose communication links with respect to the typical UE is NLoS/NLoS. Since communication can occur over both LoS and NLoS links, the conditional communication coverage probability $P_\mathrm{c}(\theta_\mathrm{c})$ can be derived by taking expectation of $P_{\mathrm{c},\rho}(\theta_\mathrm{c})$, i.e.
$P_\mathrm{c}(\theta_\mathrm{c}) = \mathbb{E}_\rho\{P_{\mathrm{c},\rho}(\theta_\mathrm{c})\}$.
Next, we can derive the expression of the moment of $P_\mathrm{c}(\theta_\mathrm{c})$.
%the probability of typical UE is associated with a Los/NLos BS is in the proof of Theorem2, did not write the explicit expression of it due to page limit.

\begin{theorem} \label{thrm:momts_com_cvrg}
	\textit{The $b$-th moment of conditional communication coverage probability $P_\mathrm{c}( \theta_\mathrm{c})$, given by
		\begin{align}\label{equ:M_b}
			M_b^\mathrm{c} = &\sum_{\rho \in \{ \mathrm{L},\mathrm{N} \}}
 \int_{0}^{\infty} \hat{f}_{\rho}(r_0) \exp\Big(-\sum_{i\in \{\mathrm{L},\mathrm{N}\}}\nonumber \\
 \int_{r_0}^{\infty}&\Big(1 - \frac{1}{(1 + \theta_\mathrm{s}\frac{ K_i}{ K_{\rho}} r_0^{\alpha_\rho}r^{-\alpha_i})^b} \Big) 2\pi \lambda_{ \mathrm{b} } rp_i(r) dr\Big) dr_0, 
		\end{align}
		with 
  % \begin{align}
  %     f_{R_0}(r_0) = 2\pi\lambda_{EQ}^*(r_0) e^{-2\pi\Lambda_{EQ}^*([0,r_0])},
  % \end{align}
  % where
  % \begin{align}
  %     \lambda_{EQ}^*(r_0) = re^{-\beta r}+[\psi^{-1}(r_0)]^2r^{-1}(1-e^{-\beta \psi^{-1}(r_0)}),
  % \end{align}
  % and 
  % \begin{align}
  %     \Lambda_{EQ}^*([0,r_0]) &= \beta^{-2}\Bigg(
  %      e^{-\beta \psi^{-1}(r_0)}(\beta(\psi^{-1}(r)) + 1)\nonumber \\
  %     &+ \frac{\beta^2}{2}[\psi^{-1}(r_0)]^2  -e^{-\beta r_0}(\beta r +1)
  %     \Bigg).
  % \end{align}
            \begin{align}
                            \label{equ:pdf_com_LOS}
                \hat{f}_\mathrm{L}(r_0) =& 2\pi \lambda_{ \mathrm{b} } r_0p_{\mathrm{L}}(r_0)\exp\Big(-2\pi\lambda_{ \mathrm{b} } \Big( - \frac{e^{-\beta r_0}(\beta r_0 + 1)}{\beta^2}  \nonumber \\
                &+ \frac{\psi^2_\mathrm{L}(r_0)}{2} + \frac{e^{-\beta \psi_\mathrm{L}(r_0)}(\beta \psi_\mathrm{L}(r_0) + 1)}{\beta^2} \Big) \Big),
                \end{align}
            \begin{align}
                            \label{equ:pdf_com_NLOS}
                \hat{f}_\mathrm{N}(r_0) = 2\pi \lambda_{ \mathrm{b} } r_0p_{\mathrm{N}}&(r_0) \exp\Big(-2\pi\lambda_{ \mathrm{b} } \Big( \frac{r_0^2}{2}+ \frac{e^{-\beta r_0}(\beta r_0 + 1)}{\beta^2}  \nonumber \\
                &- \frac{e^{-\beta \psi_\mathrm{N}(r_0)}(\beta \psi_\mathrm{N}(r_0) + 1)}{\beta^2} \Big) \Big),
            \end{align}
            where
            \begin{align}
                \psi_\mathrm{L}(r_0) = (K_\mathrm{N}/K_\mathrm{L})^{1/\alpha_\mathrm{N}}r_0^{\alpha_{\mathrm{L}}/\alpha_{\mathrm{N}}}, \\
                \psi_\mathrm{N}(r_0) = (K_\mathrm{L}/K_\mathrm{N})^{1/\alpha_\mathrm{L}}r_0^{\alpha_{\mathrm{N}}/\alpha_{\mathrm{L}}}.
            \end{align}
   	}
\end{theorem}
\begin{IEEEproof}
According to \cite[Lemma 3]{bai2014coverage}, given that the typical UE is associated with a LoS/NLoS base station, the probability density function of the distance to its serving base station, $r_0$, is given by 
\begin{align}\label{equ:pdf_com_r0}
    f_\rho(r_0) = \hat{f}_\mathrm{\rho}(r_0)/A_\rho, \rho \in \{\mathrm{L},\mathrm{N} \}.
\end{align}
where $\hat{f}_\mathrm{\rho}(r_0)$ is defined as in \eqref{equ:pdf_com_LOS} and \eqref{equ:pdf_com_NLOS} 
%and $A_\rho$ is the probability that the typical UE is associated with a LoS/NLoS base station
%
and $A_\rho$ is the probability that the typical UE is associated with a LoS/NLoS base station. Then, this may be proved in a manner similar to that of Theorem 1  by calculating the moments conditioned on $r_0$ and then de-conditioning on it using \eqref{equ:pdf_com_r0}. 
%Final result is then obtained by utilizing the law of total probability with respect to the $A_\rho$.%
\end{IEEEproof}

\subsubsection*{JCAS SIR meta distribution}
Finally, using the moments of conditional communication and sensing coverage probability, we derive the JCAS SIR meta distribution, defined as the CCDF of $P(\theta_\mathrm{c},\theta_\mathrm{s})$.

\begin{theorem} \label{thrm:meta}
	\textit{
The meta distribution of the SIR in the JCAS network under consideration is given by 
		\begin{align}\label{equ:meta}
			F( \theta_\mathrm{s}, x) = \frac{1}{2} + \frac{1}{\pi} \int_{0}^{\infty} \mathrm{Im} \Big\{ x^{ - j \omega } M_{j\omega}^\mathrm{JCAS} \Big\} \frac{d \omega}{ \omega },
		\end{align}
		 where $j = \sqrt{-1}$, $\mathrm{Im}\{ \cdot \}$ denotes the imaginary part of the input variable, and $M_b$ is the $b-th$ moment of $P(\theta_\mathrm{c},\theta_\mathrm{s})$, given by
   \begin{align}
       M_b^\mathrm{JCAS} = \frac{1}{(\lambda_\mathrm{u} +\lambda_\mathrm{s})^b}\sum^{\infty}_{m = 0} \binom{b}{m}\lambda_\mathrm{u}^{b-m}\lambda_{\mathrm{s}}^{m}M^\mathrm{c}_{b-m}M^{\mathrm{s}}_{m}.
   \end{align}
   where $M^\mathrm{c}_{b-m}$ is the $b-m$-moment of $P_\mathrm{c}(\theta_\mathrm{c})$ and $M^{\mathrm{s}}_{m}$ is the $m$-moment of $P_\mathrm{s}(\theta_\mathrm{s})$.
   	}
\end{theorem}

\begin{IEEEproof}
The $b$-th moment of $P(\theta_\mathrm{c},\theta_\mathrm{s})$ is
\begin{align}\label{equ:JCAS_moment}
    &M_b^\mathrm{JCAS} = \mathbb{E}\Big[ P(\theta_\mathrm{c},\theta_\mathrm{s})^b\Big] \nonumber \\
     &\stackrel{(a)}= \frac{1}{(\lambda_\mathrm{u} +\lambda_\mathrm{s})^b}\mathbb{E} \Big[ 
     \sum^{\infty}_{m = 0} \binom{b}{m}\lambda_\mathrm{u}^{b-m} P_\mathrm{c}(\theta_\mathrm{c})^{b-m}\lambda_{\mathrm{s}}^{m}P_\mathrm{s}(\theta_\mathrm{s})^{m} \Big] \nonumber \\
     &\stackrel{(b)}=\frac{1}{(\lambda_\mathrm{u} +\lambda_\mathrm{s})^b} 
    \sum^{\infty}_{m = 0} \binom{b}{m}\lambda_\mathrm{u}^{b-m} \mathbb{E}[P_\mathrm{c}(\theta_\mathrm{c})^{b-m}]\lambda_{\mathrm{s}}^{m}\mathbb{E}[P_\mathrm{s}(\theta_\mathrm{s})^{m}]\nonumber \\
    &= \frac{1}{(\lambda_\mathrm{u} +\lambda_\mathrm{s})^b}\sum^{\infty}_{m = 0} \binom{b}{m}\lambda_\mathrm{u}^{b-m}\lambda_{\mathrm{s}}^{m}M^\mathrm{c}_{b-m}M^{\mathrm{s}}_{m}.
\end{align}
where step ($a$) follows by binomial expansion and ($b$) follows by the independence of $P_\mathrm{c}(\theta_\mathrm{c})$ and $P_\mathrm{s}(\theta_\mathrm{s})$. 
The proof is concluded by invoking the Gil-Paleaz theorem  \cite{Gil} and then substituting \eqref{equ:JCAS_moment} into \eqref{equ:meta}. 
\end{IEEEproof}

\subsubsection*{JCAS coverage probability}
As a byproduct, we can obtain the JCAS coverage probability by computing the first moment of \eqref{equ:JCAS_moment} with repect to $\Phi_\mathrm{b}$, given as:
\begin{align}
    \label{equ:M1}
    &M_1^\mathrm{JCAS} = \frac{\lambda_\mathrm{u}M_1^\mathrm{c} + \lambda_\mathrm{s}M_1^\mathrm{s}}{\lambda_\mathrm{u} + \lambda_\mathrm{s}}
\end{align}
which is consistent with the definition in \cite{olson2022coverage}. Such a metric reflects the joint fraction of UEs and SOs whose coverage conditions are satisfied. 

\section{Numerical Results} \label{Sec:SimNum_Results}
We now provide numerical results to validate our analysis and evaluate how the JCAS SIR meta distribution is affected by the SIR thresholds and by the BS density. 
We generate 100 PPP realizations for the locations of BSs, UEs, and SOs. 
Within the Voronoi cell formed by the BSs, UEs, and SOs are situated in a square area measuring 1,000 m in width on each side. Once constructed, the topology remains unchanged, but the fading realizations of communications and sensing across each link are recalculated across 1,000 time periods. 
Then, we collect the statistics of communications and sensing to compute the conditional JSAC coverage probability of each realization. 
We adopt the following parameters unless otherwise noted: $\alpha_\mathrm{L} = 2$, $\alpha_\mathrm{N} = 3.2$, $K_\mathrm{L} = -75.96$ dB, $K_\mathrm{N} = -90.96$ dB, $\beta = 1/140$, $\lambda_\mathrm{b} = 10^{-4} $~m$^{-2}$, and $\lambda_\mathrm{u} =\lambda_\mathrm{s} = 10^{-3}$~m$^{-2}$.
     
\begin{figure}
\centering
\begin{subfigure}{0.9\columnwidth}
    \includegraphics[width=1\columnwidth]{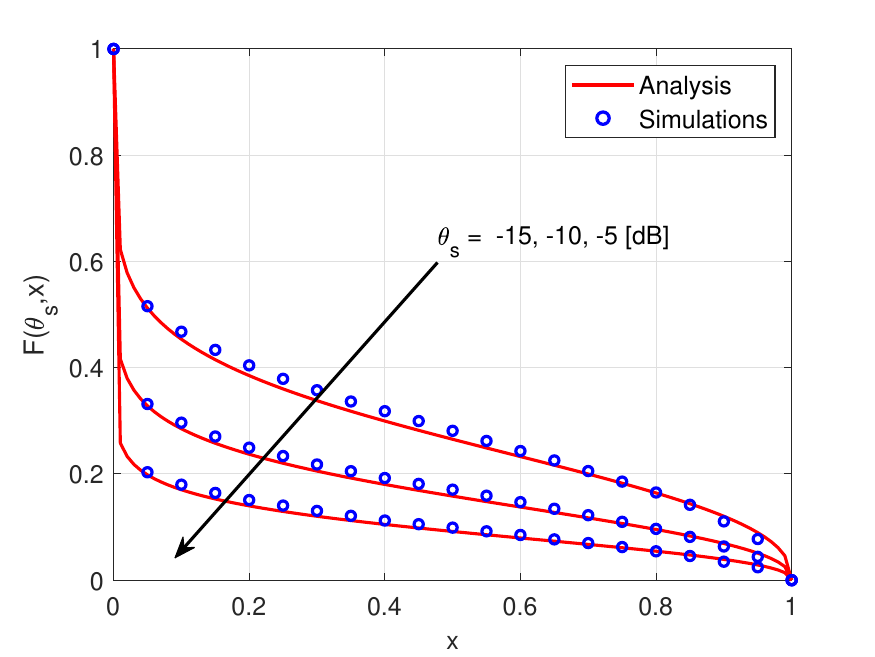}
    \caption{Sensing SIR meta distribution for $\theta_{\mathrm{s}}$ = -15, -10, -5 dB.}
    \label{fig:first}
\end{subfigure}
 \hfill
 \begin{subfigure}{0.9\columnwidth}
     \includegraphics[width=1\columnwidth]{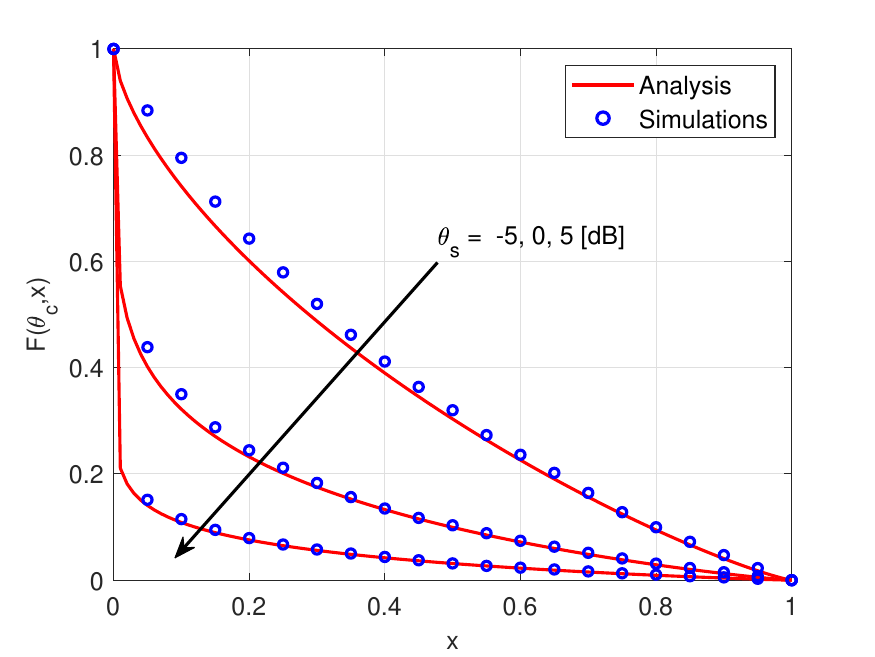}
     \caption{Communication SIR meta distribution for $\theta_{\mathrm{c}}$ =  -5, 0, 5 dB.}
     \label{fig:second}
\end{subfigure}
% \hfill
% \begin{subfigure}{0.9\columnwidth}
%     \includegraphics[width=0.9\columnwidth]{Figures/CCDF3.eps}
%     %\caption{Meta distribution for $\theta_{\mathrm{c}},\theta_{\mathrm{s}}$ = (0,-5), (5,0), (10, 5) dB.}
%     \label{fig:third}
% \end{subfigure}
\caption{SIR meta distribution (Theorem~\ref{thrm:momts_Sen_cvrg}, Theorem~\ref{thrm:momts_com_cvrg}) and simulations, as a function of the reliability threshold (x-axis) and for different SIR thresholds ($\theta_\mathrm{c},\theta_\mathrm{s}$).}
\label{fig:figures}
\end{figure}

% \begin{figure}[t!]
%   \centering{}
%     {\includegraphics[width=0.9\columnwidth]{Figures/SenMetaVsDensityAndBlockage.eps}}
%   \caption{Sensing SIR meta distribution as a function of the reliability threshold (x-axis) under different blockage parameter and BSs density }
%   \label{fig:density}
% \end{figure}

Fig.~\ref{fig:figures} compares the simulated CCDF of the conditional sensing and communication coverage probability (black circles) to the analysis in Theorem~\ref{thrm:momts_com_cvrg} and Theorem~\ref{thrm:momts_Sen_cvrg}, under various pairs of communication and sensing thresholds.
% The close match of all three validates both our exact mathematical derivations in Theorem~\ref{lma:momts_JCAS_cvrg} and the proposed approximation based on Corollary~\ref{cor:aprx_momts}.
% 
The JCAS SIR meta distribution in Fig.~\ref{fig:figures} provides a fine-grained evaluation of the network performance in terms of both communication and sensing. For instance, the Fig.~\ref{fig:figures}(a) shows that for $\theta_{\mathrm{s}} = -15$\,dB and $-5$ dB, respectively, setting a reliability threshold of 0.4 on the x-axis corresponds to values of about 0.3 and 0.1 on the y-axis. This indicates that 30\% of the SOs in this network can achieve sensing SIRs of at least $-15$\,dB with a 40\% reliability, but this fraction drops to 10\% when both SIR thresholds are raised to $-5$\,dB.

\begin{figure}[t!]
  \centering{}
    {\includegraphics[width=0.9\columnwidth]{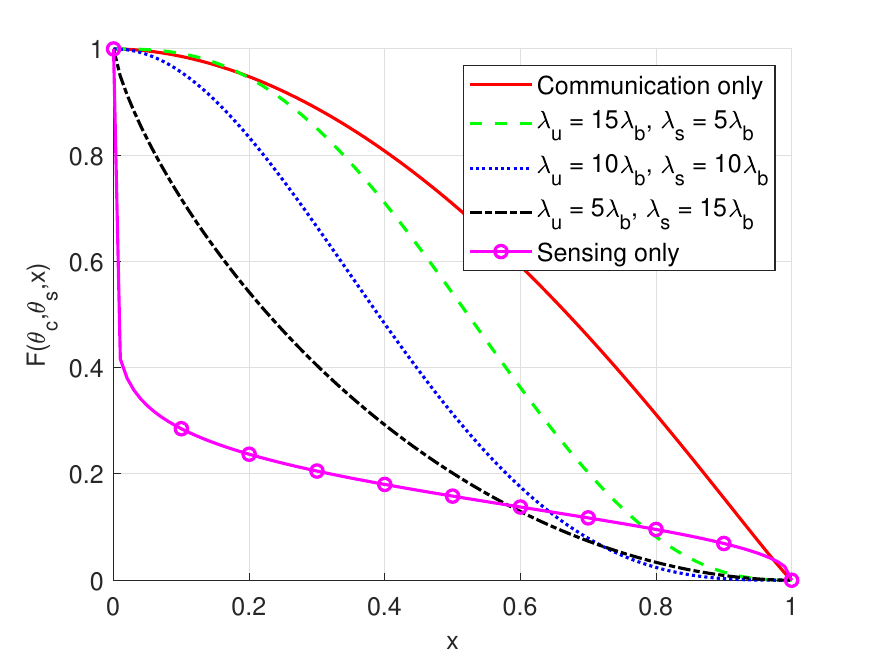}}
  \caption{JCAS SIR meta distribution as a function of the reliability threshold under different UEs and SOs densities. SIR detection threshold is set as $\theta_{\mathrm{c}} = \theta_{\mathrm{s}} = -10$ dB.}
  \label{fig:JCASMeta}
\end{figure}

Fig.~\ref{fig:JCASMeta} plots the JCAS SIR meta distribution in Theorem~\ref{thrm:meta} under various UEs and SOs densities. 
We observe that under the same SIR detection threshold $\theta_{\mathrm{c}} = \theta_{\mathrm{s}}$, scenarios with a larger ratio $\lambda_{\mathrm{u}} / \lambda_{\mathrm{s}}$ exhibit a higher JCAS SIR coverage. This is due to the fact that, unlike communication, sensing experiences double path loss and can only take place in LoS conditions. Therefore, increasing the proportions of scheduled communication transmissions with respect to sensing transmissions results in a higher coverage probability.
%
%We observe that under the same SIR detection threshold, networks with larger UE densities have wider JCAS SIR coverage, and thus, increasing the proportions of UEs can improve the JCAS network performance. The reasons come from the fact that sensing signals experience double path loss, and sensing is LoS only while communication may occur over either LoS or NLoS links.
%
It is notable that for large values of reliability threshold, the meta distribution of sensing may exceed that of communication. This anomalous behavior occurs when the serving BS is unrealistically close to the typical UE/SO, resulting in higher signal strength for sensing under the adopted power law path loss model.

% Fig.\ref{fig:density} illustrates the impact of blockage patameters $\beta$ and BSs density $\Phi_\mathrm{b}$ on the meta distribution of sensing SIR when the SIR threshold $\theta_\mathrm{s}$ is $-10$ dB. We observe that the meta distribution increases when decreasing blockage parameter and increasing BSs density. This is because decreasing blockage parameter

\begin{figure}[t!]
  \centering{}
    {\includegraphics[width=0.9\columnwidth]{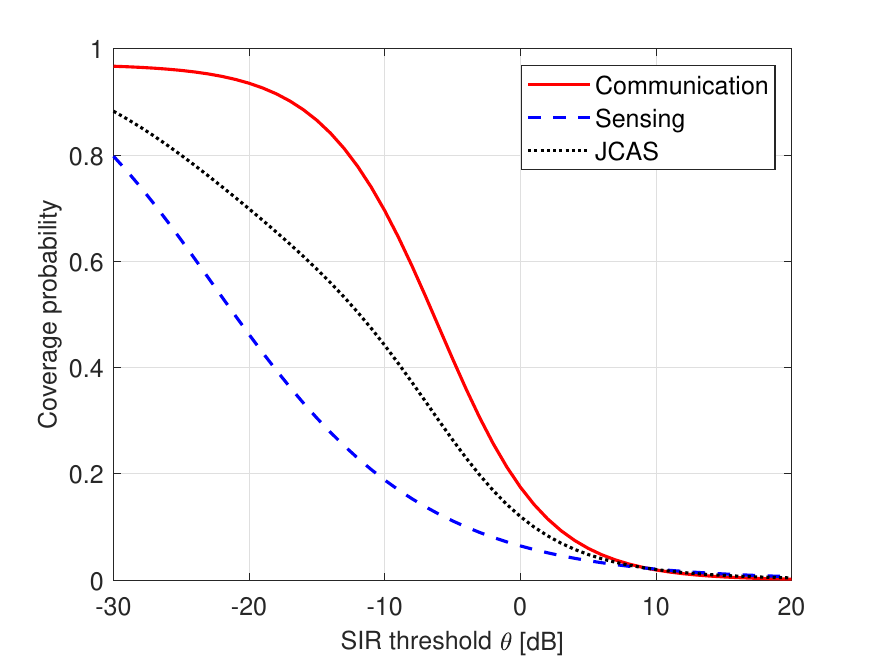}}
  \caption{Coverage probability of communication only, sensing only, and JCAS versus SIR threshold.}
  \label{fig:coverage}
\end{figure}
% \begin{figure}[t!]
%   \centering{}
%     {\includegraphics[width=0.95\columnwidth]{Figures/SenMetaVsDensityAndBlockage.eps}}
%   \caption{Sensing SIR meta distribution as a function of the reliability threshold (x-axis) and under different blockage parameter and BSs density }
%   \label{fig:density}
% \end{figure}

In Fig.~\ref{fig:coverage}, we plot the coverage probabilities of communication, sensing, and JCAS vs. the decoding thresholds. 
% While for convenience of plotting the two SIR thresholds are equally set as $\theta_\mathrm{c} = \theta_\mathrm{s}$, our framework allows to compute the coverage probability for any combination of values. 
Fig.~\ref{fig:coverage} shows that compared to sensing, the coverage probability of communication is more sensitive to changes in the SIR threshold. This indicates that the sensing performance exhibits more variability across SOs than the communication performance across UEs. 
% Besides, under same SIR threshold, sensing coverage probability is lower compared with the communication case. This is because sensing signal experiences double path loss and sensing is LoS only while communication may occurs over either LoS or NLoS links. 
%interference has a more significant effect on the quality of communications than sensing.

% Fig.~\ref{fig:density} depicts the sensing coverage probability as a function of the BS deployment density for three values of the SIR threshold. In line with the findings in \cite{olson2022coverage}, the figure shows that the sensing coverage probability increases with the BS deployment density. This is in contrast to what occurs to communications under the adopted single-slope pathloss assumption, where the coverage probability is independent of the BS density, as captured in (\ref{equ:com_cvrg}).

%========================================%
%                Conclusion
%========================================%
\section{Conclusion} \label{Sec:Conclusion}

In this paper, we developed a framework to assess the performance of JCAS in wireless networks, utilizing stochastic geometry as a key analytical tool. 
Our approach involved deriving mathematical expressions for the conditional JCAS coverage probability and its distribution, known as the SIR meta distribution. 
% Given the computational complexity of these expressions, we also presented practical, tight approximations for simpler numerical evaluations.
Our theoretical models, validated through simulations, capture the impact of network deployment density on the JCAS SIR performance. 
Unlike previous studies, our analysis allows to quantify individual user or radar performance within network realizations.

While this work primarily focused on the impact of network deployment density on the JCAS SIR meta distribution, we note that this distribution is also influenced by the propagation environment model and the specific deployment scenario. Future research directions include incorporating realistic BS deployment, channel, and antenna models, and exploring dynamics in the temporal domain, such as data traffic for communication \cite{YanQuePoo:19TCOM} and status updates for sensing \cite{YanXuWan:19}.
%========================================%
%            Appendix A
%========================================%

\section*{Appendix: Proof of Theorem \ref{thrm:momts_Sen_cvrg}} \label{thm1}
The $b$-th moment of $P_\mathrm{s}(\theta_\mathrm{s} )$ can be expressed as:	 
\begin{align}\label{equ:mb_half}
	 	M_b^\mathrm{s} =& \mathbb{E}_{r_0} \Bigg[
   \underbrace{\mathbb{E} \Big[ 
   % \prod_{ k \neq 0 } \Big(\frac{1}{1 + \theta_\mathrm{s}  \frac{ L_1(\Vert X_k-X_0 \Vert)}{L_2(\Vert X_0 \Vert)}}\Big)^b
      \prod_{i \in \{ \mathrm{L},\mathrm{N} \} } \prod_{ X_k \in \Phi^i_{ \mathrm{b} \setminus X_0 } } \!\!\!\! \Bigg(  \frac{1}{1 + \frac{ \theta_\mathrm{s} K_i\Vert X_0 \Vert ^{\alpha_\mathrm{L}}}{ K_{\mathrm{L}}\Vert X_k -X_0 \Vert^{\alpha_i}} } \Bigg)^b \!\!
   \mid r_0 
   \Big] }_{M^\mathrm{s}_{b\mid r_0}} \Bigg]
   \nonumber \\
   &\times \bigg( 1-\exp\Big(\frac{-2\pi\lambda_{\mathrm{b}}}{\beta^2}\Big) \bigg)^b,
\end{align}
Then, one must determine the moment of conditional coverage probability of sensing conditioned on the distance between the serving BS and the origin, $r_0 = \Vert X_0 \Vert$.

Lemma 2 of \cite{olson2022coverage} states that the point process of the distance of interfering BSs with regard to serving BS conditioned on $r_0$ is a PPP on $\mathbb{R_+}$ with the following intensity function, provided as $\Pi^0_B = \{ \Vert X_k-X_0 \Vert: X_k \in \Phi_\mathrm{b}^{!X_0}\}$:
            \begin{align}
            \lambda^0_{\mathrm{b}}(r;r_0)=2\lambda_{\mathrm{b}} r\Big(\pi-J(r,r_0)\mathbbm{1}\{r \le 2r_0\}\Big),
            \end{align}
            where $J(r,r_0)$ is defined as in \eqref{equ:Jr}.
            %\begin{align}
                %J(r,r_0)=\int_{\frac{r}{2r_0}}^1 (1-u^2)^{-1/2} \exp\Big( -\beta \sqrt{r^2-2rr_0u+r_0^2}\Big)du.
            %\end{align}
            
Then, via the probability generating
functional (PGFL) of PPP and applying independent thinning theorem to $\Pi^0_\mathrm{b}$ with respect to the LoS/NLoS status, the conditional moment of the sensing  coverage probability can be derived as
\begin{align}\label{equ:Msbr0}
		%M^s_{b|r_0}  \nonumber \\ 
		M^\mathrm{s}_{b|r_0}  
  % &=  \exp \Big( -\int_{0}^{\infty}\Big[1-\frac{1}{\big(1+\theta_{\mathrm{s}} \frac{L_1(r)}{L_2(r_0)} \big)^b}\Big]\lambda_\mathrm{b}^0(r;r_0)dr\Big)\nonumber\\
        & = \exp\Big(-\sum_{i\in \{\mathrm{L},\mathrm{N}\}}\int_{0}^{\infty}\Big(1 - \frac{1}{(1 + \theta_\mathrm{s}\frac{ K_i}{ K_{\mathrm{L}}} r_0^{2\alpha_L}r^{-\alpha_i})^b} \Big) \nonumber \\
        &\quad \quad \quad \quad \quad \quad \quad \quad \quad \quad \times p_i(r)\lambda^0_{\mathrm{b}}(r;r_0)dr\Big).
 \end{align}
% where steps (a) follows by applying Independent Thinning Theorem to $\Pi^0_\mathrm{b}$ with repect to the LoS/NLoS status.
% , resulting in two independent process: $\Pi^0_\mathrm{L}$ and $\Pi^0_\mathrm{N}$, whose intensity measures are $p_\mathrm{L}(r)\lambda_\mathrm{b}^0(r;r_0)$ and $p_\mathrm{N}(r)\lambda_\mathrm{b}^0(r;r_0)$.

Given the typical user observes at least one LoS base station, the conditional probability density function of its distance to the nearest LOS base station is given by \cite[Lemma 1]{bai2014coverage}:
\begin{align}\label{equ:pdf_r0}
    f_\mathrm{L}(x) = \frac{2\pi \lambda_{\mathrm{b}} x p_{\mathrm{L}}(x) \exp\Big(\frac{-2\pi\lambda_{\mathrm{b}}}{\beta^2}(1-e^{-\beta x}(\beta x+ 1)) \Big)}{1-\exp(-2\pi\lambda_{\mathrm{b}}/\beta^2)}.
\end{align}
Then, substituting \eqref{equ:Msbr0} into \eqref{equ:mb_half} and de-conditioning on $r_0$ using \eqref{equ:pdf_r0}, we obtain the expression of the $b$-th moment given in \eqref{equ:M_b}.

\acresetall

%\balance
\bibliographystyle{IEEEtran}
\bibliography{bib/StringDefinitions,bib/IEEEabrv,bib/ATB_SALOHA}

\begin{thebibliography}{10}
\providecommand{\url}[1]{#1}
\csname url@samestyle\endcsname
\providecommand{\newblock}{\relax}
\providecommand{\bibinfo}[2]{#2}
\providecommand{\BIBentrySTDinterwordspacing}{\spaceskip=0pt\relax}
\providecommand{\BIBentryALTinterwordstretchfactor}{4}
\providecommand{\BIBentryALTinterwordspacing}{\spaceskip=\fontdimen2\font plus
\BIBentryALTinterwordstretchfactor\fontdimen3\font minus \fontdimen4\font\relax}
\providecommand{\BIBforeignlanguage}[2]{{%
\expandafter\ifx\csname l@#1\endcsname\relax
\typeout{** WARNING: IEEEtran.bst: No hyphenation pattern has been}%
\typeout{** loaded for the language `#1'. Using the pattern for}%
\typeout{** the default language instead.}%
\else
\language=\csname l@#1\endcsname
\fi
#2}}
\providecommand{\BIBdecl}{\relax}
\BIBdecl

\bibitem{TarKhaWon:20WCMAG}
F.~Tariq, M.~R. Khandaker, K.-K. Wong, M.~A. Imran, M.~Bennis, and M.~Debbah, ``A speculative study on {6G},'' \emph{IEEE Wireless Commun.}, vol.~27, no.~4, pp. 118--125, Aug. 2020.

\bibitem{OugGerPol2023}
E.~Oughton, G.~Geraci, M.~Polese, and V.~Shah, ``Prospective evaluation of next generation wireless broadband technologies: {6G} versus {Wi-Fi} 7/8,'' \emph{Available at SSRN: https://ssrn.com/abstract=4528119}, 2023.

\bibitem{DanAmiShi:20NE}
S.~Dang, O.~Amin, B.~Shihada, and M.-S. Alouini, ``What should {6G} be?'' \emph{Nature Electronics}, vol.~3, no.~1, pp. 20--29, 2020.

\bibitem{liu2020joint}
F.~Liu, C.~Masouros, A.~P. Petropulu, H.~Griffiths, and L.~Hanzo, ``Joint radar and communication design: Applications, state-of-the-art, and the road ahead,'' \emph{IEEE Trans. Commun.}, vol.~68, no.~6, pp. 3834--3862, 2020.

\bibitem{zhang2021enabling}
J.~A. Zhang, M.~L. Rahman, K.~Wu, X.~Huang, Y.~J. Guo, S.~Chen, and J.~Yuan, ``Enabling joint communication and radar sensing in mobile networks—{A} survey,'' \emph{IEEE Commun. Surv. Tutor.}, vol.~24, no.~1, pp. 306--345, 2021.

\bibitem{ren2020performance}
P.~Ren, A.~Munari, and M.~Petrova, ``Performance analysis of a time-sharing joint radar-communications network,'' in \emph{2020 International Conference on Computing, Networking and Communications (ICNC)}.\hskip 1em plus 0.5em minus 0.4em\relax IEEE, 2020, pp. 908--913.

\bibitem{salem2022rethinking}
A.~Salem, C.~Masouros, F.~Liu, and D.~L{\'o}pez-P{\'e}rez, ``Rethinking dense cells for integrated sensing and communications: A stochastic geometric view,'' \emph{Available as ArXiv:2212.12942}, 2022.

\bibitem{olson2022coverage}
N.~R. Olson, J.~G. Andrews, and R.~W. Heath~Jr, ``Coverage and capacity of joint communication and sensing in wireless networks,'' \emph{Available as ArXiv:2210.02289}, 2022.

\bibitem{ghozlani2021stochastic}
D.~Ghozlani, A.~Omri, S.~Bouallegue, H.~Chamkhia, and R.~Bouallegue, ``Stochastic geometry-based analysis of joint radar and communication-enabled cooperative detection systems,'' in \emph{Proc. IEEE WiMob}, 2021, pp. 325--330.

\bibitem{ma2022performance}
H.~Ma, Z.~Wei, Z.~Li, F.~Ning, X.~Chen, and Z.~Feng, ``Performance of cooperative detection in joint communication-sensing vehicular network: A data analytic and stochastic geometry approach,'' \emph{IEEE Trans. Veh. Technol.}, vol.~72, no.~3, pp. 3848--3863, 2022.

\bibitem{haenggi2016meta}
M.~Haenggi, ``The meta distribution of the {SIR} in {Poisson} bipolar and cellular networks,'' \emph{IEEE Trans. Wireless Commun.}, vol.~15, no.~4, pp. 2577--2589, Apr. 2016.

\bibitem{Hae:21CL-1}
M.~Haenggi, ``Meta distributions—{P}art 1: Definition and examples,'' \emph{IEEE Commun. Lett.}, vol.~25, no.~7, pp. 2089--2093, Jul. 2021.

\bibitem{Hae:21CL-2}
M.~Haenggi, ``Meta distributions—{P}art 2: Properties and interpretations,'' \emph{IEEE Commun. Lett.}, vol.~25, no.~7, pp. 2094--2098, Jul. 2021.

\bibitem{ghatak2022radar}
G.~Ghatak, S.~S. Kalamkar, and Y.~Gupta, ``Radar detection in vehicular networks: Fine-grained analysis and optimal channel access,'' \emph{IEEE Trans. Veh. Technol.}, vol.~71, no.~6, pp. 6671--6681, 2022.

\bibitem{ram2022optimization}
S.~S. Ram, S.~Singhal, and G.~Ghatak, ``Optimization of network throughput of joint radar communication system using stochastic geometry,'' \emph{Frontiers in Signal Processing}, vol.~2, p. 835743, 2022.

\bibitem{bai2014analysis}
T.~Bai, R.~Vaze, and R.~W. Heath, ``Analysis of blockage effects on urban cellular networks,'' \emph{IEEE Trans. Wireless Commun.}, vol.~13, no.~9, pp. 5070--5083, 2014.

\bibitem{Zuyev2006}
A.~Baddeley, P.~Gregori, J.~Mateu, R.~Stoica, and D.~Stoyan, \emph{Case Studies in Spatial Point Process Modeling. Lecture Notes in Statistics}.\hskip 1em plus 0.5em minus 0.4em\relax Springer, 2006.

\bibitem{richards2005fundamentals}
M.~A. Richards, \emph{Fundamentals of radar signal processing}.\hskip 1em plus 0.5em minus 0.4em\relax Mcgraw-hill New York, 2005, vol.~1.

\bibitem{shnidman2003expanded}
D.~Shnidman, ``Expanded swerling target models,'' \emph{IEEE Trans. Aerosp. Electron. Syst.}, vol.~39, no.~3, pp. 1059--1069, 2003.

\bibitem{bai2014coverage}
T.~Bai and R.~W. Heath, ``Coverage and rate analysis for millimeter-wave cellular networks,'' \emph{IEEE Trans. Wireless Commun}, vol.~14, no.~2, pp. 1100--1114, 2014.

\bibitem{Gil}
J.~Gil-Pelaez, ``Note on the inversion theorem,'' \emph{Biometrika}, vol.~38, no. 3-4, pp. 481--482, Dec. 1951.

\bibitem{YanQuePoo:19TCOM}
H.~H. Yang, T.~Q.~S. Quek, and H.~V. Poor, ``A unified framework for {SINR} analysis in {Poisson} networks with traffic dynamics,'' \emph{IEEE Trans. Commun.}, vol.~69, no.~1, pp. 326--339, Jan. 2021.

\bibitem{YanXuWan:19}
H.~H. Yang, C.~Xu, X.~Wang, D.~Feng, and T.~Q.~S. Quek, ``Understanding age of information in large-scale wireless networks,'' \emph{IEEE Trans. Wireless Commun.}, vol.~20, no.~5, pp. 3196--3210, May 2021.

\end{thebibliography}

\end{document}